\documentclass[12pt]{article}
\usepackage{amssymb,latexsym,amsfonts,amsmath,color,framed,esint,mathrsfs}
\usepackage[colorlinks]{hyperref}

\begin{document}
\title{The helicity and vorticity of liquid crystal flows}
\author{Fran\c{c}ois Gay-Balmaz$^1$ and Cesare Tronci$^2$
\\
\vspace{-.5cm}
\\
{\footnotesize $^1$ \it Control and dynamical systems, California Institute of Technology}\\
{\footnotesize $^2$ \it Section de Math\'ematiques, \'Ecole
Polytechnique F\'ed\'erale de Lausanne, Switzerland}
}
\date{}

\maketitle

\begin{abstract}
We present explicit expressions of the helicity conservation in
nematic liquid crystal flows, for both the Ericksen-Leslie and
Landau-de Gennes theories. This is done by using a
minimal coupling argument that leads to an Euler-like equation for a
modified vorticity involving both velocity and structure fields
(e.g. director and alignment tensor). This equation for the modified
vorticity shares many relevant properties with ideal fluid dynamics
and it allows for vortex filament configurations as well as point
vortices in 2D. We extend all these results to particles of
arbitrary shape by considering systems with fully broken rotational
symmetry.
\end{abstract}



\bigskip

\section{Introduction} Several studies on nematic liquid crystal
flows have shown high velocity gradients and led to the conclusion
that the coupling between the velocity
$\boldsymbol{u}(\mathbf{x},t)$ and structure fields is a fundamental
feature of liquid crystal dynamics \cite{ReyDenn02}. This conclusion
has been reached from different viewpoints and by using different
theories, such as the celebrated Ericksen-Leslie (EL) and the
Landau-de Gennes (LdG) theories
\cite{Chonoetal,Tothetal,TaoFeng,BlSvetal}. Evidence of high
velocity gradients also emerged \cite{KuKaDe} by using kinetic
approaches based on the Doi model \cite{DoEd1988}. The essential
difference between EL and LdG theories resides in the choice of the
order parameter: while EL theory for rod-like molecules considers
the dynamics of the director field $\mathbf{n}(\mathbf{x},t)$ and it
is successful in the description of low molar-mass nematics, the LdG
theory generalizes to variable molecule shapes by considering a
traceless symmetric tensor field ${\sf Q}(\mathbf{x},t)$. In the
presence of high disclination densities the LdG theory is more reliable,
since molecules may easily undergo phase transitions (e.g. from
uniaxial to biaxial order) that are naturally incorporated in the
theory. However, the dynamical LdG theory is not completely
established and different versions are available in the literature
\cite{BeEd1994,Lubensky2003,QiSh1998}. Here, we shall adopt the
formulation of Qian and Sheng \cite{QiSh1998}, which will be simply
referred to as LdG theory.

This paper considers both EL and LdG theories and it shows how the
strong interplay between velocity and order parameter field reflects
naturally in the helicity conservation for nematics. In this paper, the term ``helicity'' stands for the hydrodynamic helicity and \emph{not} the helicity of the single liquid crystal molecule. The helicity
conservation for incompressible liquid crystal flows arises from the
simple velocity transformation
$\boldsymbol{u}\to\boldsymbol{\mathcal{C}}=\boldsymbol{u}+\mathbf{J}$,
where the vector $\mathbf{J}$ depends only on the order parameter
field. The covariant vector $\boldsymbol{\mathcal{C}}$ is the total circulation (momentum per unit mass of fluid) and $\mathbf{J}$ is the circulation associated with entrainment of fluid due to its local interaction with the nematic order parameter field.  We shall show how this change of velocity  variable takes the  equation for
the ordinary vorticity
$\boldsymbol\omega=\nabla\times\boldsymbol{u}$ into an Euler-like
equation for the modified vorticity
$\overline{\boldsymbol\omega}=\nabla\times\boldsymbol{\mathcal{C}}$,
thereby extending many properties of ordinary ideal fluids to
nematic liquid crystals. The helicity is then given by
\[
\mathcal{H}=\int \left( \overline{\boldsymbol{ \omega }} \cdot
\boldsymbol{\mathcal{C}} \right) \,\mathrm{d}^3\mathbf{x},
\]
and this quantity naturally extends the usual expression $\int
\left(\boldsymbol{ \omega } \cdot \boldsymbol{ u} \right)\,
\mathrm{d}^3\mathbf{x}$ for the helicity of three dimensional ideal
flows. We recall that in Hamiltonian fluid dynamics the conservation of  the hydrodynamic helicity is strictly
associated to the Hamiltonian structure of the equations and holds
for any Hamiltonian. This point will be further developed in the last part of the paper, where all the results will be derived directly from the Hamiltonian structure of the liquid crystal equations, see \cite{GayBalmazTronci}. Invariant functions like the helicity are called Casimir and are of
fundamental importance for the study of nonlinear stability. For two
dimensional flows, such invariant quantities are given by
\begin{equation}\label{Casimir2D}
\int \Phi(\overline{ { \omega }})\ \mathrm{d}^2\mathbf{x},
\end{equation}
where $\Phi$ is an arbitrary smooth function. As we shall see, the same circulation concept leading to hydrodynamic helicity applies quite generally in complex fluid theory and is related to an analogy between complex fluids and non-Abelian Yang-Mills fluid plasmas \cite{GiHoKu1983,HoKu1988}. 

In addition to helicity conservation, we present the
existence of vortex-like configurations for the modified vorticity $
\overline{\boldsymbol\omega}$. Vortex structures are well known to
arise in superfluid flows and their behavior is often reminiscent of
disclination lines in liquid crystals. However, here we shall
consider vortices that are characterized by a combination of
velocity and structures fields.  After extending these
results to fluids with molecules of arbitrary shapes, the end of this paper discusses the geometric basis of the present treatment.

\bigskip

\section{Director formulation} In the context of EL theory,
disclinations are singularities of the director field and thus their
dynamics is related to the evolution of the gradient
$\nabla\mathbf{n}$. This relation has been encoded by Eringen
\cite{Eringen} in the \emph{wryness tensor}
\begin{equation}\label{GammaDef}
\gamma_\text{\tiny
EL}=\mathbf{n}\times\nabla\mathbf{n}\,,\quad
\text{ or }\quad
\left(\gamma_\text{\tiny
EL}\right)^a_i=\varepsilon^{abc\,}{n}_b\times\partial_i{n}_c\,,
\end{equation}
which identifies the amount by which the director field rotates
under an infinitesimal displacement $\mathrm{d}\mathbf{x}$. Thus,
the EL wryness tensor $\gamma_\text{\tiny EL}$ determines
the spatial rotational strain \cite{Ho2002}. In this paper we shall investigate the role of the EL wryness tensor in helicity conservation and vorticity dynamics in the EL theory. 
For this purpose, we ignore dissipation and concentrate on nonlinearity. This simplifies the resulting formulas. We also restrict to incompressible flows to ignore ordinary fluid thermodynamics. 

Upon
denoting by $J$ the microinertia constant \cite{Eringen}, we
introduce the angular momentum variable
\begin{equation}\label{SigmaDef}
\boldsymbol\sigma_\text{\tiny EL}=J\mathbf{n}\times D_t{\mathbf{n}}
\end{equation}
 that is associated to the director precession. While $D_t{\mathbf{n}}=\partial_t{\mathbf{n}}+\boldsymbol{u}\cdot\nabla{\mathbf{n}}$ denotes material time-derivative, the
spatial derivatives of the director field will be denoted
equivalently by $\partial \mathbf{n}/\partial
x^i=\partial_i\mathbf{n}=\mathbf{n}_{,i}$ depending on convenience.
Upon using Einstein's summation convention, one can
express the Ericksen-Leslie equations as
\cite{Ho2002,GayBRatiu,GayBalmazTronci}
\begin{align}\label{EL1}
\partial_t \boldsymbol{u}+\boldsymbol{u}\!\cdot\!\nabla \boldsymbol{u} &=-\partial_i\!\left(\nabla\mathbf{n}^T\!\cdot\!\frac{\partial F}{\partial\mathbf{n}_{,i}}\right)-\nabla p\,,\quad \  \nabla\cdot\boldsymbol{u} =0
\\
\partial_t{\boldsymbol{\sigma}}_\text{\tiny EL}+\boldsymbol{u} \cdot\!\nabla\boldsymbol{\sigma}_\text{\tiny EL}&=\mathbf{h}\times\mathbf{n}\,,\hspace{.7cm} \partial_t{\mathbf{n}}+\boldsymbol{u} \cdot\!\nabla\mathbf{n}=J^{-1}\boldsymbol{\sigma}_\text{\tiny EL}\times\mathbf{n}
\label{EL2}
\end{align}
where $p$ is the hydrodynamic pressure by which incompressibility is imposed and $\mathbf{h}$ is the thermodynamics derivative representing the First Law response in energy to changes in the director field
\[
\mathbf{h}:=\frac{\partial
F}{\partial\mathbf{n}}-\partial_{i\!}\left(\frac{\partial
F}{\partial\mathbf{n}_{,i}}\right)\,.
\]
The quantity $F$ is taken to be the Oseen-Z\"ocher-Frank free energy
\begin{equation}\label{standard_energy}
F=K_1(\operatorname{div}\mathbf{n})^2+K_2(\mathbf{n}\cdot\nabla\times\mathbf{n})^2+K_3|\mathbf{n}\times\nabla\times\mathbf{n}|^2\,.
\end{equation}
Of course, this choice is not a limitation, because our considerations apply to a generic form of $F$. For example, effects
of external electric and magnetic fields may be taken into account with easy modifications. The Ericksen-Leslie fluid equations follow immediately from equations \eqref{EL1} and \eqref{EL2}, as shown in \cite{GayBRatiu}.

In Ericksen-Leslie nematodynamics, the quantity $\boldsymbol{\sigma}_\text{\tiny
EL}\cdot\gamma_\text{\tiny EL}$ denotes the vector of momentum  per unit mass, with components ${\left(\boldsymbol{\sigma}_\text{\tiny
EL}\cdot\gamma_\text{\tiny EL}\right)_i}={\left({\sigma}_\text{\tiny
EL}\right)_a\left(\gamma_\text{\tiny EL}\right)_i^a}$. We consider the vector $\boldsymbol{\mathcal{C}}_\text{\tiny
EL}$ defined as the sum
\begin{equation}\label{Cdef}
\boldsymbol{\mathcal{C}}_\text{\tiny
EL}:=\boldsymbol{u}+\boldsymbol{\sigma}_\text{\tiny
EL}\cdot{\gamma}_\text{\tiny
EL}=\boldsymbol{u}+\boldsymbol{\sigma}_\text{\tiny
EL}\cdot\mathbf{n}\times\nabla\mathbf{n}
\,,\end{equation}
reminiscent of the minimal coupling formula in electromagnetic gauge theory.
We observe the following equation of motion \cite{GayBalmazTronci} (see appendix \ref{appendix1}):
\footnote{
{ The same idea has been applied in superfluid plasmas, that is, in superfluid solutions 
whose charged condensates are coupled electromagnetically \cite{HoKu1987}. However, }
the similarity with gauge theory does not end with the electromagnetic analogy. The equation corresponding to (\ref{Cequation}) also follows by inspection for a Yang-Mills fluid plasma (chromohydrodynamics, cf. \cite{GiHoKu1983}) either from equation (2.35) or (2.49) of \cite{HoKu1988}.  By this observation, chromohydrodynamics acquires a circulation theorem and the theory of complex fluids inherits an analogy with Yang-Mills fluid plasma, first noticed in \cite{Ho2002}. This minimal coupling argument requires the wryness tensor $\gamma_\text{\tiny EL}$ to be a connection one-form: although this is not the case for the expression $\mathbf{n}\times\nabla\mathbf{n}$, a connection one-form can be obtained by the addition of  terms parallel to $\mathbf{n}$. {By good fortune, these extra terms make no contribution in (\ref{Cdef}) because $\boldsymbol{\sigma}_\text{\tiny EL}\cdot\mathbf{n}=0$.}  }
\begin{equation}\label{Cequation}
\partial_t\boldsymbol{\mathcal{C}}_\text{\tiny
EL}-\boldsymbol{u}\times\nabla\times\boldsymbol{\mathcal{C}}_\text{\tiny
EL} =-\nabla\!\left(
\phi+\boldsymbol{u}\cdot\boldsymbol{\mathcal{C}}_\text{\tiny
EL}\right) \,,
\end{equation}
where
\begin{equation}
\phi=p+F-\frac12\left|\mathbf{u}\right|^2-\frac1{2J}\left|\boldsymbol{\sigma}_\text{\tiny
EL}\right|^2 \,.
\end{equation}
At this point, taking the curl of equation \eqref{Cequation} yields the Euler-like equation
\begin{equation}\label{EulerForNematics}
\partial_t\overline{\boldsymbol{\omega}}_\text{\tiny
EL}+\nabla\times\left(\boldsymbol{u}\times\overline{\boldsymbol{\omega}}_\text{\tiny
EL}\right)=0\,
\end{equation}
for the modified
vorticity
\begin{equation}\label{NewVort}
\overline{\boldsymbol{\omega}}_\text{\tiny
EL}:=\nabla\times\boldsymbol{\mathcal{C}}_\text{\tiny
EL}=\boldsymbol{\omega}+\nabla\times\left(\boldsymbol{\sigma}_\text{\tiny
EL}\cdot{\gamma}_\text{\tiny EL}\right)
\,.
\end{equation}
Notice that the velocity $\boldsymbol{u}$ can be expressed as
\begin{equation}\label{untangled-u}
\boldsymbol{u}=-\nabla\times\boldsymbol\psi=-\nabla\times\Delta^{-1}\overline{\boldsymbol\omega}_\text{\tiny
EL}+\boldsymbol{\sigma}_\text{\tiny
EL}\cdot{\gamma}_\text{\tiny EL}+\nabla\varphi \,,
\end{equation}
where $\boldsymbol\psi=\Delta^{-1}\boldsymbol{\omega}$ denotes the
velocity potential, which is given by the convolution of the
vorticity $\boldsymbol{\omega}$ with the Green's function of the
Laplace operator (analogously for
$\Delta^{-1}\overline{\boldsymbol\omega}_\text{\tiny EL}$). Here the
pressure-like quantity $\varphi$ is a scalar function arising from
the term
$
{\nabla\times\nabla\times\Delta^{-1}\left(\boldsymbol{\sigma}_\text{\tiny
EL}\cdot{\gamma}_\text{\tiny
EL}\right)}={{\boldsymbol{\sigma}_\text{\tiny
EL}\cdot{\gamma}_\text{\tiny EL}}+\nabla\varphi}$
 and
whose only role is to keep the velocity $\boldsymbol{u}$ divergence
free, so that 
$
{\nabla\cdot(\boldsymbol{\sigma}_\text{\tiny
EL}\cdot{{\gamma}_\text{\tiny EL}})}=-\Delta\varphi
$.
The
relation \eqref{untangled-u} can be inserted into equations
\eqref{EL2} so to express the EL equations in terms of the modified
vorticity $\overline{\boldsymbol\omega}_\text{\tiny EL}$. An
explicit expression of the quantity $\boldsymbol{\sigma}_\text{\tiny
EL}\cdot{\gamma}_\text{\tiny EL}$ arises from the
definitions \eqref{GammaDef} and \eqref{SigmaDef}:
$
{\boldsymbol{\sigma}_\text{\tiny
EL}\cdot{\gamma}_\text{\tiny EL}}={J\,\nabla\mathbf{n}\cdot
D_t\mathbf{n}}
$.

At this point we recognize that equation \eqref{EulerForNematics}
possesses all the usual properties of Euler's equation. For example, Ertel's commuting
relation
\begin{equation}\label{ertel}
\big[{D_{t\,}},\,\overline{\boldsymbol{\omega}}_\text{\tiny
EL}\cdot\nabla\big]\,\alpha={D_t}\!\left(\overline{\boldsymbol{\omega}}\cdot\nabla\alpha\right)-\overline{\boldsymbol{\omega}}\cdot\nabla\!
\left({D_{t\,}}\alpha\right)=0
\end{equation}
follows easily by direct verification, for any scalar function $\alpha(\mathbf{x},t)$. Moreover, one has
the following Kelvin circulation theorem \cite{GayBalmazTronci}
\begin{equation}\label{kelvin}
\frac{d}{dt}\oint_{\Gamma(t)}
\!\big(\boldsymbol{u}+\boldsymbol{\sigma}_\text{\tiny
EL}\cdot\gamma_\text{\tiny
EL}\big)\cdot\mathrm{d}\mathbf{x}=0\,,
\end{equation}
where the line integral is calculated on a loop $\Gamma(t)$ moving
with velocity $\boldsymbol{u}$. Also, conservation of the helicity \cite{GayBalmazTronci}
\[
\mathcal{H}_\text{\tiny
EL}=\!\int\overline{\boldsymbol{\omega}}_\text{\tiny
EL}\cdot\boldsymbol{\mathcal{C}}_\text{\tiny EL}\
\mathrm{d}^3\mathbf{x}
=\!\int\!\left(\boldsymbol{u}+\boldsymbol{\sigma}_\text{\tiny
EL}\cdot\gamma_\text{\tiny
EL}\right)\cdot\nabla\times\left(\boldsymbol{u}+\boldsymbol{\sigma}_\text{\tiny
EL}\cdot\gamma_\text{\tiny
EL}\right)
\mathrm{d}^3\mathbf{x}
\]
follows from the relation
\begin{equation}\label{h-equation}
\partial_t\left(\boldsymbol{\mathcal{C}}_\text{\tiny
EL}\cdot\overline{\boldsymbol{\omega}}_\text{\tiny
EL}\right)+\nabla\cdot\big(\left(\boldsymbol{\mathcal{C}}_\text{\tiny
EL}\cdot\overline{\boldsymbol{\omega}}_\text{\tiny
EL}\right)\boldsymbol{u}\big)=-\,\overline{\boldsymbol{\omega}}_\text{\tiny
EL}\cdot\nabla \phi
\,,
\end{equation}
which is obtained by using equations \eqref{EulerForNematics} and
\eqref{Cequation}. Integrating equation \eqref{h-equation} over the
fluid volume yields
\begin{equation}\label{h-equation-int}
\frac{d}{dt}\mathcal{H}_\text{\tiny EL}=-\varoiint_S\, \phi \
\overline{\boldsymbol{\omega}}_\text{\tiny
EL}\cdot\mathrm{d}\mathbf{S}-\varoiint_S\left(\boldsymbol{\mathcal{C}}_\text{\tiny
EL}\cdot\overline{\boldsymbol{\omega}}_\text{\tiny
EL}\right)\boldsymbol{u}\cdot\mathrm{d}\mathbf{S}
\end{equation}
where $S$ is the surface determined by the fluid boundary.
Consequently, the right hand side of equation \eqref{h-equation-int}
vanishes when $\overline{\boldsymbol{\omega}}_\text{\tiny EL}$ and
$\boldsymbol{u}$ are both tangent to the boundary, thereby producing
conservation of $\mathcal{H}_\text{\tiny EL}$. Remarkably, one can show that helicity conservation persists for
\emph{any} free energy $F(\mathbf{n},\nabla\mathbf{n})$, that is the
 helicity $\mathcal{H}_\text{\tiny EL}$ is a Casimir
for EL dynamics, see \cite{GayBalmazTronci}. 
At this point, a question about boundary conditions arises: while the condition of velocity tangent to the boundary is the usual condition in hydrodynamics, the condition 
\begin{equation}\label{b-cond}
\nabla\times\left(\boldsymbol{u}+\boldsymbol{\sigma}_\text{\tiny
EL}\cdot\mathbf{n}\times\nabla\mathbf{n}\right)\cdot\mathrm{d}\mathbf{S}=0
\end{equation}
 emerges here for the first time. Upon denoting $\boldsymbol\pi=\boldsymbol\sigma_\text{\tiny EL}\times\mathbf{n}$, one has ${\nabla\times\left(\boldsymbol{u}+\boldsymbol{\sigma}_\text{\tiny
EL}\cdot\mathbf{n}\times\nabla\mathbf{n}\right)}={\boldsymbol\omega+\nabla\pi_a\times\nabla n_a}$, so that the boundary condition \eqref{b-cond} reads as ${\nabla\pi_a\times\nabla n_a\cdot\mathrm{d}\mathbf{S}}={-\,\boldsymbol\omega\cdot\mathrm{d}\mathbf{S}}$. This relation evidently differs from the usual ``anchoring'' boundary conditions (see e.g. \cite{Stewart}) that are widely used in the literature and for which the director alignment at the surface is insensitive to the flow. Indeed, the physical relevance of the boundary condition \eqref{b-cond} resides in the fact that it involves \emph{both} fluid and field variables, contrarily to other commonly available boundary conditions. The complete physical justification of \eqref{b-cond}, however, requires more study in the future.

One of the most relevant consequences of equation
\eqref{EulerForNematics} is the
existence of singular vortex-like configurations in Ericksen-Leslie nematodynamics. In two dimensions,
equation \eqref{EulerForNematics} has the usual point vortex
solution
\[
\overline{\omega}_\text{\tiny
EL}(x,y,t)=\sum_{i=1}^Nw_i\,\delta(x-X_i(t))\,\delta(y-Y_i(t)) \,,
\]
where $(X_i,Y_i)$ are canonically conjugate variables with respect to
the Hamiltonian $\psi=\sum(\Delta^{-1}\omega)(X_i,Y_i)$. Here $\psi$ is
 the potential of the velocity $\boldsymbol{u}$, satisfying
EL equation \eqref{EL1}. Upon using relation \eqref{NewVort} and suppressing the EL label for convenience, one expresses the Hamiltonian as
\begin{align*}\nonumber
\psi(X_i,Y_i,\boldsymbol{\sigma},\mathbf{n})=
&-\frac1{4\pi}\sum_{h}\left(\sum_kw_h\,w_k\log|(X_h-X_k,Y_h-Y_k)|\right.
\\
&+\left.w_h\!\int\!\big\{\sigma_a,\gamma^a\big\}(x',y')\,\log|(x'-X_h,y'-Y_h)|\,\mathrm{d}x'\,\mathrm{d}y'\right)
,
\end{align*}
where $|(x,y)|=\sqrt{x^2+y^2}$ and $\{\cdot,\cdot\}$ denotes the canonical Poisson bracket in $(x,y)$ coordinates, arising from the 2D relation
$
{\nabla\times\left(\boldsymbol{\sigma}\cdot{\gamma}\right)}
=
{\nabla\times\left(\nabla\mathbf{n}\cdot\boldsymbol{\pi}\right)}
=
{\big\{\pi_a,\,{n}_{a}\big\}}
$, where $\boldsymbol\pi= \boldsymbol\sigma\times\mathbf{n}$.

 Other
vortex-like structures are also allowed by equation
\eqref{EulerForNematics}, e.g. vortex cores and patches. In three
dimensions, the vortex filament
\[
\overline{\boldsymbol{\omega}}_\text{\tiny
EL}(\mathbf{x},t)=\int\frac{\partial\mathbf{R}(s,t)}{\partial s}\
\delta(\mathbf{x}-\mathbf{R}(s,t))\ \mathrm{d}s
\]
is also a solution of \eqref{EulerForNematics}, with
$\partial_t\mathbf{R}=\boldsymbol{u}(\mathbf{R},t)$. The existence of these vortex structures (including vortex sheets) rise natural stability questions concerning possible equilibrium vortex configurations. Instead of pursuing this direction, which will be the subject of our future work, the next sections will show how all the above observations also hold in the LdG theory and for fluid molecules of arbitrary shape.

To conclude this section, we emphasize that in all the above discussion the
velocity and the structure fields are strongly coupled together.
Indeed, the singular vortex structures only exist for the
vorticity $\overline{\boldsymbol{\omega}}_\text{\tiny
EL}=\nabla\times\boldsymbol{\mathcal{C}}_\text{\tiny EL}$, while
there is no way for the ordinary vorticity
$\boldsymbol{\omega}=\nabla\times\boldsymbol{u}$ or the `director
vorticity' $\nabla\times(\boldsymbol{\sigma}_\text{\tiny
EL}\cdot\boldsymbol{\gamma}_\text{\tiny EL})$ to be singular. This
strong interplay between the macro- and micromotion is the same that emerges in many of the experiments and simulations
reviewed in \cite{ReyDenn02}.

\section{The alignment tensor} In the preceding section, we
investigated the hydrodynamics of a uniaxial nematic liquid crystal.
At this point, it is natural to argue that in the presence of
disclinations the molecules can change the configuration of their
order parameter (e.g. from uniaxial to biaxial) and the EL equations
cannot be used as a faithful model, which is rather given by the LdG
theory based on the alignment tensor $\sf Q$. Several dynamical
fluid models for the evolution of the alignment tensor $\mathsf{Q}$
are found in the literature \cite{BeEd1994,Lubensky2003,QiSh1998}.
In this paper we shall show how the ideal Qiang-Sheng (QS) model
\cite{QiSh1998} for the LdG tensor order parameter also allows for
helicity conservation, in analogy to EL theory.

The ideal QS model reads as
\begin{align}\label{QS1}
&\partial_t \boldsymbol{u}+\boldsymbol{u}\!\cdot\!\nabla
\boldsymbol{u} = -\partial_l\!\left(\frac{\partial
\mathcal{F}}{\partial\mathsf{Q}_{ij\,,l}}\nabla\mathsf{Q}_{ij}\right)-\nabla
p\,,\ \ \nabla\cdot\boldsymbol{u} =0
\\ \label{QS2}
&\partial_t\mathsf{Q}+\boldsymbol{u} \cdot\!\nabla
\mathsf{Q}=J^{-1}\, \mathsf{P}
\\
&\partial_t\mathsf{P}+\boldsymbol{u} \cdot\!\nabla\mathsf{P}=
-\frac{\partial\mathcal{F}}{\partial
\mathsf{Q}}+\partial_i\frac{\partial\mathcal{F}}{\partial
\mathsf{Q}_{,i}}-\lambda \mathbf{I}
 \label{QS3}
\end{align}
where  $\mathsf{P}$ is conjugate to $\mathsf{Q}$ and $\mathbf{I}$ is
the identity matrix, while $\lambda$ is a Lagrange multiplier
arising from the condition $\operatorname{Tr}\mathsf{Q}=0$. Here the
free energy $\mathcal{F}(\mathsf{Q},\nabla\mathsf{Q})$ contains the
Landau-de Gennes free energy \cite{deGePr1993} as well as
interaction terms with external fields. Notice that the molecular
field ${\partial\mathcal{F}/\partial
\mathsf{Q}}-{\partial_i(\partial\mathcal{F}/\partial \mathsf{Q}_{,i})}$
is always symmetric, so that $\mathsf{P}$ is also symmetric at all
times.

The circulation vector $\boldsymbol{\mathcal{C}}_\text{\tiny
QS}$ for the above system is defined by
\[
\boldsymbol{\mathcal{C}}_\text{\tiny
QS}:=\boldsymbol{u}+\mathsf{P}_{ij}\nabla\mathsf{Q}_{ij} \,.
\]
Indeed, a direct verification shows that the above vector satisfies
equation \eqref{Cequation}, that is (cf appendix \ref{appendix2})
\begin{equation}\label{Ceq-Qtensor}
\partial_t\boldsymbol{\mathcal{C}}_\text{\tiny QS}+\nabla\left(\boldsymbol{u}\cdot\boldsymbol{\mathcal{C}}_\text{\tiny QS}\right)-\boldsymbol{u}\times\nabla\times\boldsymbol{\mathcal{C}}_\text{\tiny QS}
=-\nabla \phi
\,,
\end{equation}
with
\begin{equation}
\phi=p+\mathcal{F}-\frac12\left|\boldsymbol{u}\right|^2-\frac1{2J}\mathsf{P}_{ij}\mathsf{P}_{ij}
\,.
\end{equation}
Thus, the Euler-like equation
\begin{equation}\label{LdGEuler}
\partial_t\overline{\boldsymbol{\omega}}_\text{\tiny QS}+\nabla\times\left(\boldsymbol{u}\times\overline{\boldsymbol{\omega}}_\text{\tiny QS}\right)=0
\end{equation}
holds for the modified vorticity
$\overline{\boldsymbol{\omega}}_\text{\tiny
QS}=\nabla\times\boldsymbol{\mathcal{C}}_\text{\tiny QS}$. The
circulation theorem
\begin{equation}
\frac{d}{dt}\oint_{\Gamma(t)}
\!\left(\boldsymbol{u}+\mathsf{P}_{ij}\nabla\mathsf{Q}_{ij}\right)\cdot\mathrm{d}\mathbf{x}=0\,,
\end{equation}
and the helicity conservation (for
$\overline{\boldsymbol{\omega}}_\text{\tiny QS}$ and
$\boldsymbol{u}$ both tangent to the boundary)
\[
\frac{d}{dt}\int\boldsymbol{\mathcal{C}}_\text{\tiny
QS}\cdot\overline{\boldsymbol{\omega}}_\text{\tiny QS}\
\mathrm{d}^3\mathbf{x}=0
\]
are a natural consequence of the Euler-like equation
\eqref{LdGEuler} for the QS model of LdG theory. Again Ertel's
commutation relation \eqref{ertel} for
$\overline{\boldsymbol{\omega}}_\text{\tiny QS}$ follows easily from
equation \eqref{LdGEuler}. Moreover, vortex structures similar to those
appearing in EL theory also exist in the LdG formulation.

At this point, one can ask about other LdG formulations and in particular one wonders whether the latter also exhibit vortex structures and conservation of hydrodynamic helicity. Among the LdG formulations of liquid crystal dynamics, the one by Beris and Edwards \cite{BeEd1994} is probably among the most common, although it  is not known to possess helicity conservation. In particular, this theory treats the order parameter as a ``conformation tensor field'', so that the symmetric matrix $\sf Q$ is replaced by a symmetric tensor field on physical space. This deep geometric difference is probably responsible for the absence of the hydrodynamic helicity in the Beris-Edwards formulation. 
In this sense, the peculiarity of the QS model for the LdG tensor dynamics resides in exhibiting helicity
conservation and its associated vorticity dynamics. These quantities both involve coupling between
velocity and structure fields. The next section shows how this is actually a
situation common to all fluid systems exhibiting rotational symmetry
breaking.

\section{Completely broken symmetries}
 The tensor order parameter $\sf Q$ arises as usual from
the broken rotational symmetry that is typical of liquid crystal
materials. When this rotational symmetry is fully broken, one needs
to account for the dynamics of the whole particle orientation, which
is determined by an orthogonal matrix $O$ (such that $O^{-1}=O^T$).
This is a situation occurring, for example, in spin glass dynamics
\cite{Fischer,DzVo,IsKoPe}. Eringen's wryness tensor (here denoted by $\boldsymbol{\kappa}$) is written in terms of $O$ as \cite{Eringen}
\begin{equation}\label{NewWryness}
\kappa^s_i=\frac12\,\varepsilon^{mns} \,O_{mk}\, \partial_i O_{nk}\,
=\frac12\,\varepsilon^{mns} \, \partial_i O_{nk}\, O_{km}^{-1}
\,,\quad\text{ (sum over repeated indexes})
\end{equation}
(In this section we suppress labels such as EL or QS, in order to
better adapt to the tensor index notation.) Although orthogonal
matrices are difficult to work with analytically and the use of
quaternions could be preferable, the correspondence between
quaternions and rotation matrices is not unique. Thus, following
Eringen's work \cite{Eringen}, we identify molecule orientations
with orthogonal matrices.

It is the purpose of this section to show how equation
\eqref{EulerForNematics} is not peculiar of nematic liquid crystals.
Rather, equations of this form are peculiar of all systems with
 broken rotational symmetries. In particular, equation
\eqref{EulerForNematics} also holds in the case of complete symmetry
breaking, for a vorticity variable $\overline{\boldsymbol\omega}$
depending on the fluid velocity $\boldsymbol{u}$, on the full
particle orientation $O$ and on the angular momentum vector
\begin{equation}\label{NewSigma}
\sigma_r=\varepsilon_{rmn} \,O_{mk}\, \Psi_{nk}\, =\varepsilon_{rmn}
\, \Psi_{nk}\, O_{km} ^{-1}
\end{equation}
where $\Psi$ is the variable conjugate to $O$.
 In the well-known spin glass
theory of Halperin and Saslow \cite{HaSa}, the rotation matrix is
small and thus it is replaced by its infinitesimal rotation angle
$\boldsymbol\theta$, where $\exp(\boldsymbol\theta)=O$. Then,
$\boldsymbol\theta$ and $\boldsymbol\sigma$ become canonically
conjugate variables, as shown in \cite{Fischer}. Here we
consider the whole matrix $O$ to account for arbitrary rotations.

In the case of fully broken symmetry, the (incompressible) equations
of motion can be written for an arbitrary energy  density
$\mathcal{E}(\boldsymbol{\sigma},O)$ as \cite{IsKoPe}
\begin{align}\label{SG1}
&\partial_t \boldsymbol{u}+\boldsymbol{u}\!\cdot\!\nabla
\boldsymbol{u} =-
\sigma_r\nabla\frac{\partial\mathcal{E}}{\partial\sigma_r}+\frac{\partial\mathcal{E}}{\partial
O_{mn}}\nabla O_{mn}
 -\nabla p
\\ \label{SG2}
&\partial_t{\sigma}_r+\boldsymbol{u}
\cdot\!\nabla\sigma_r=\varepsilon_{r\!ji}\left(\sigma_i\frac{\partial\mathcal{E}}{\partial
\sigma_j} - O_{ih}\frac{\partial\mathcal{E}}{\partial O_{jh}}\right)
\\
&\partial_tO_{mn}+\boldsymbol{u} \cdot\!\nabla
O_{mn}=-\varepsilon_{mkj}\frac{\partial\mathcal{E}}{\partial
\sigma_j}O_{kn} \label{SG3}
\end{align}
Notice that the validity of the above set of equations is completely general. Indeed, the above system is derived in \cite{IsKoPe} in a general fashion, under the only hypothesis that the broken symmetry group is $SO(3)$. More general broken symmetries can be certainly treated in the same way, although this paper focuses only on rotational symmetries.

 In this context, equation \eqref{Cdef} generalizes immediately by considering the wryness tensor in
\eqref{NewWryness}. Thus, the new circulation vector is defined by
\begin{equation}\label{NewC}
\boldsymbol{\mathcal{C}}:=\boldsymbol{u}+\boldsymbol{\sigma}\cdot{\kappa}=\boldsymbol{u}+\frac12\,\varepsilon_{mns}\,\sigma_s\,
O_{mk}\nabla O_{nk}\,,
\end{equation}
where $\boldsymbol{u}$, $\boldsymbol{\sigma}$ and $O$ satisfy
equations \eqref{SG1}, \eqref{SG2} and \eqref{SG3} .

At this point it is natural to ask whether the new
$\boldsymbol{\mathcal{C}}$ satisfies equation \eqref{Cequation}.
Remarkably, a positive answer again arises from a direct calculation by
using the ordinary properties of the Levi-Civita symbol. One obtains (see appendix \ref{appendix3})
\begin{equation}\label{Ceq-Omodel}
\partial_t\boldsymbol{\mathcal{C}}+\nabla\left(\boldsymbol{u}\cdot\boldsymbol{\mathcal{C}}\right)-\boldsymbol{u}\times\nabla\times\boldsymbol{\mathcal{C}}
=-\nabla \left(p-\frac12\left|\boldsymbol{u}\right|^2\right) \,.
\end{equation}
Consequently, the Euler-like equation \eqref{EulerForNematics} holds
also in this case when the rotational symmetry is completely broken.
Explicitly one writes
\begin{equation}\label{NewCequation}
\partial_t (\nabla\times\boldsymbol{\mathcal{C}})+
\nabla\times\big(\boldsymbol{u}\times\nabla\times\boldsymbol{\mathcal{C}}\big)=0
\,.
\end{equation}
In turn, equations \eqref{NewCequation} and \eqref{NewC} imply the
circulation law
\begin{equation}\label{NewCirculation}
\frac{d}{dt}\oint_{\Gamma(t)}
\!\left(\boldsymbol{u}+\frac12\,\varepsilon_{mns}\,\sigma_s\,
O_{mk}\nabla O_{nk}\right)\cdot\mathrm{d}\mathbf{x}=0\,,
\end{equation}
and the helicity conservation
\[
\frac{d}{dt}\int_V\boldsymbol{\mathcal{C}}\cdot\overline{\boldsymbol{\omega}}\
\mathrm{d}^3\mathbf{x}=0\,,
\]
with
$\overline{\boldsymbol{\omega}}=\nabla\times\boldsymbol{\mathcal{C}}$.
Thus, the existence of vortex configurations is independent of the
type of symmetry breaking characterizing the fluid. Therefore,
such vortices may exist in liquid crystals independently of the
choice of order parameter. However, one should also emphasize that
the energy conserving assumption may fail in several situations and
one would then be forced to consider viscosity effects. Moreover, polymeric liquid crystals do not seem to fit easily into the present framework; rather their description requires other liquid crystal theories such as
the celebrated Doi theory \cite{DoEd1988}.

\section{Geometric origin of the helicity invariant}
In the previous sections, the helicity and vorticity of various systems with broken symmetry have been presented. However, the explicit formulation of these results still lack some more justification that can be found in the deep geometric nature of these systems, as it was emphasized in \cite{GayBalmazTronci}. This section aims to give a brief overview of the geometric setting of the liquid crystal equations that eventually leads to the explicit formulation of their helicity and vorticity. This will show how these quantities can be found without any of the calculation presented in the Appendix, by simply relying on geometric symmetry concepts. The reader is also addressed to \cite{MaRa,Ho2002}.

As our starting point, we write the total Poisson bracket for a general (incompressible) fluid system with broken symmetry, involving an order parameter space $M$.
 In this case, the dynamical variables consist of the fluid momentum $\mathbf{m(x)}$, the order parameter state $\mathcal{Q}(\mathbf{x})\in M$ and its conjugate variable $\mathcal{P}(\mathbf{x})$, so that $(\mathcal{Q}(\mathbf{x}),\mathcal{P}(\mathbf{x}))\in T^*M$. For simplicity, we restrict to consider the case when $M$ is a matrix vector space. The total Poisson bracket reads as
\begin{align}
\{F,G\}=&\int\mathbf{m}\cdot\left[\frac{\delta F}{\delta \mathbf{m}},\frac{\delta G}{\delta \mathbf{m}}\right]\mathrm{d}^3x+\int\operatorname{Tr}\!\left(\left(\frac{\delta F}{\delta \mathcal{Q}}\right)^{\!T}\frac{\delta G}{\delta \mathcal{P}}-\left(\frac{\delta F}{\delta \mathcal{P}}\right)^{\!T}\frac{\delta G}{\delta \mathcal{Q}}\right)\mathrm{d}^3x
\nonumber
\\
&
+
\left\langle \frac{\delta
F}{\delta (\mathcal{Q},\mathcal{P})}\,,\pounds_{\frac{\delta G}{\delta \mathbf{m}}} \left(\mathcal{Q},\mathcal{P}\right) \right\rangle
-
\left\langle\frac{\delta
G}{\delta (\mathcal{Q},\mathcal{P})}\,, \pounds_{\frac{\delta F}{\delta \mathbf{m}}}\left(\mathcal{Q},\mathcal{P}\right) \right\rangle,
\label{PB-general}
\end{align}
where the angle bracket denotes the pairing
\begin{align}
\left\langle\frac{\delta
G}{\delta (\mathcal{Q},\mathcal{P})}\,, \pounds_{\frac{\delta F}{\delta \mathbf{m}}}\left(O,\mathcal{P}\right) \right\rangle=
\int\operatorname{Tr}\!\left(\left(\frac{\delta
G}{\delta \mathcal{Q}}\right)^{\!T\!}\pounds_{\frac{\delta F}{\delta \mathbf{m}}}\mathcal{Q}+\left(\frac{\delta
G}{\delta \mathcal{P}}\right)^{\!T\!}\pounds_{\frac{\delta F}{\delta \mathbf{m}}}\mathcal{P}\right).
\end{align}
The above bracket is derived from the relabeling symmetry that characterizes all fluid systems.
In particular, this bracket characterizes all Hamiltonian fluid systems with broken symmetry. The relabeling symmetry carried by the fluid emerges mathematically as an invariance property of the Hamiltonian functional $\mathscr{H}:T^{*\!}\operatorname{Diff}(\Bbb{R}^3)\times T^*C^\infty(\Bbb{R}^3,M)\to\Bbb{R}$ under the diffeomorphism group $\operatorname{Diff}(\Bbb{R}^3)$ of smooth invertible maps. Here the notation $C^\infty(\Bbb{R}^3,M)$ stands for the space of $M$-valued scalar functions, i.e. the space of order parameter fields. The reduction process induces a reduced Hamiltonian ${H:\mathfrak{X}^*(\Bbb{R}^3)\times T^*C^\infty(\Bbb{R}^3,M)\to\Bbb{R}}$, where $\mathfrak{X}^*(\Bbb{R}^3)$ denotes the space of differential one-forms, i.e. the space of fluid momentum vectors $\mathbf{m(x)}$. This process leading to the reduced Hamiltonian $H=H(\mathbf{m},\mathcal{Q},\mathcal{P})$ is widely explained in \cite{MaRa,Ho2002,KrMa1987}.

Each term in the above Poisson bracket possesses a precise geometric meaning. While the first term coincides with the Poisson bracket for ordinary fluids, the second term is the canonical bracket for the order parameter field $\mathcal{Q}(\mathbf{x})$ and its conjugated momentum $\mathcal{P}(\mathbf{x})$. Moreover, the whole second line contains the two terms arising from the action of the relabeling symmetry group $\operatorname{Diff}(\Bbb{R}^3)$ on the canonical order parameter variables $(\mathcal{Q}(\mathbf{x}),\mathcal{P}(\mathbf{x}))$. Poisson brackets of this form were  applied in different contexts, from electromagnetic charged fluids \cite{Ho1986,Ho1987} to superfluid dynamics \cite{HoKu1982}, {and even to superfluid plasmas \cite{HoKu1987}.}

At this point, upon following the Hamiltonian version of Noether's theorem (see \cite{MaRa}), 
one can construct the total momentum
\[
\boldsymbol{\mathcal{C}}=\mathbf{m}+\mathbf{J}(\mathcal{Q},\mathcal{P})
\]
where $\mathbf{J}(\mathcal{Q},\mathcal{P})$ is the (cotangent-lift) momentum map of components
\[
{J}_i(\mathcal{Q},\mathcal{P})=\operatorname{Tr}\!\left(\mathcal{P}^T\partial_i\mathcal{Q}\right).
\]
The geometric meaning of this momentum shift by a momentum map is best explained in \cite{KrMa1987}.
In Lie derivative notation, the dynamics of the circulation quantity reads as
\begin{equation}\label{C-general}
\left(\frac{\partial}{\partial t}+\pounds_{\textstyle\frac{\delta H}{\delta \mathbf{m}}}\right)\boldsymbol{\mathcal{C}}=-\nabla\phi
\end{equation}
thereby yielding Noether's conservation relation
\[
\frac{d}{dt}\oint_{\Gamma(t)} \boldsymbol{\mathcal{C}}\cdot\operatorname{d}\mathbf{x}=0
\]
which then arises naturally as the circulation conservation determined by the relabeling symmetry of the system. The explicit proof of circulation theorems of this kind can be found in many works in geometric fluid dynamics; see \cite{HoMaRa1998} for a modern reference. After recalling that for incompressible flows $\mathbf{m}=\boldsymbol{u}$, it is easy to recognize that replacing $M$ by the space $\operatorname{Sym}_0(3)$ of traceless symmetric matrices transforms the relation \eqref{C-general} exactly into the relation \eqref{Ceq-Qtensor}, which then produces the results in Section 3.  Moreover, the corresponding vorticity relation for $\overline{\boldsymbol\omega}=\mathbf{d} \boldsymbol{\mathcal{C}}$ is easily obtained by taking the exterior differential of equation \eqref{C-general} and recalling that this operation commutes with Lie derivative. Then, one obtains $\left(\partial_t+\pounds_{\boldsymbol{u}}\right)\mathbf{d} \boldsymbol{\mathcal{C}}=0$. The form of the helicity is also easily derived from the above arguments, upon recalling an old result in \cite{KrMa1987}. In particular, if $\mathcal{H}({\bf m})$ denotes ordinary Euler's helicity, then $\mathcal{H}(\boldsymbol{\mathcal{C}})$ is a Casimir invariant of the Poisson bracket \eqref{PB-general}. Notice that  all the above relations hold for an arbitrary manifold $M$ other than a matrix space. This only requires using the appropriate pairing between vectors and co-vectors.

So far, we only used the cotangent-lift momentum map, which can be found for all the cases when the dynamics involve conjugate variables in a cotangent bundle $T^*M$. However, this does not appear to be the case for the discussion in Section 4, where $M=SO(3)$ and $\mathcal{Q}=O$. This apparent contradiction is easily solved by noticing that 
\[
\operatorname{Tr}\!\left(\mathcal{P}^T\,\partial_iO\right)
=
\operatorname{Tr}\!\left((\mathcal{P}O^{-1})^T\partial_iO\,O^{-1}\right).
\]
Then, upon denoting $\hat{\sigma}=\mathcal{P}O^{-1}$ and $\hat{\kappa_i}=\partial_iO\,O^{-1}$, the usual isomorphism between antisymmetric matrices in the Lie algebra $\mathfrak{so}(3)$ and vectors in $\mathbb{R}^3$ yields the term $\boldsymbol\sigma\cdot\kappa$ in the circulation quantity \eqref{NewC}. Then, upon repeating the same steps as above, the momentum map $\operatorname{Tr}\!\left(\mathcal{P}^T\partial_iO\right)=\boldsymbol\sigma\cdot\kappa$ returns exactly the same results as in Section 4.

The case of nematic liquid crystals treated in Section 2 can be also obtained by a direct computation, upon setting $M=S^2/\Bbb{Z}_2$, which is the director space. Upon denoting $\boldsymbol{\pi}=JD_t{\mathbf{n}}$ the corresponding conjugate variable, it is easy to see that $\nabla\mathbf{n}\cdot\boldsymbol{\pi}=\boldsymbol\sigma\cdot\gamma_\text{\tiny
EL}$. However, the geometric meaning of this simple step requires more basis that can be found in \cite{GayBalmazTronci}, where this last relation is justified by Lagrangian reduction.

At this point, it is clear that the above arguments ensure the results in this paper without any need for further discussion. Nevertheless, the Appendix gives explicit proofs that can be followed without previous knowledge in geometric mechanics. 

\section{Conclusions} This paper provided explicit expressions for
the helicity conservation in liquid crystals, in both EL and LdG
theories. This conservation arises from an Euler-like equation that
allows for singular vortex structures in any dimension. Some of the
ideal fluid properties were extended to liquid crystal flows, e.g.
Ertel's commutation relation. These results were also shown to hold
for molecules of arbitrary shapes, by considering fully broken
rotational symmetries occurring in some spin glass dynamics. All of the results were eventually justified by geometric symmetry arguments. 

The energy-Casimir method can then be applied to study nonlinear stability properties of these systems, see \cite{HoMaRaWe1985} for several examples of how this method applies to many types of fluids. This can be used, for example, to explore the coupled macro- and micro-motion of the stationary (generalized Beltrami) solutions. While the 3D  stability analysis is limited by the fact that the helicity is the only Casimir invariant, the 2D case is much reacher because the whole family \eqref{Casimir2D} of Casimir invariants becomes available.

One more remark concerns physically observable effects. More particularly, one wonders how conservation of total circulation causes observable effects. Even more, one would like to observe these effects
in a particular experiment. A simple technique that could be used to this purpose is the use of external electric fields that drive the order-parameter variables, thereby generating fluid circulation by conservation of the total circulation. Then, if one applies an external field to a trivial motionless liquid crystal, the director alignment caused by the field would result in the generation of fluid motion. 

Other physical questions also arise about the nature of vortex solutions, which evidently represent much more that simply disclinations dragged around by a smooth flow. It is possible that these solutions share many analogies with superfluid vortices in He$^3$-A, whose order parameter space is again the whole group $SO(3)$. However, the nature of these singular solutions is left open for future investigations, together with their stability properties.

\paragraph{Acknowledgements.} We are indebted with Darryl Holm
for his keen remarks about the relation between the modified
vorticity and helicity conservation in Ericksen-Leslie theory. Some
of this work was carried out while visiting him at Imperial College
London.

\appendix
\section{Appendix}

\subsection{Derivation of equations \eqref{Cequation} and \eqref{EulerForNematics}\label{appendix1}}
Upon using the notation $\pounds_{\boldsymbol{u}}$ for the Lie
derivative with respect to the velocity vector field
$\boldsymbol{u}$ \cite{MaRa}, we can rewrite equations
\eqref{EL1}-\eqref{EL2} as
\begin{align}\label{EL1-bis}
\left(\frac{\partial}{\partial t}
+\pounds_{\boldsymbol{u}}\right)\boldsymbol{u}
&=\nabla\mathbf{n}\cdot\mathbf{h}
-\nabla\left(p+F-\frac12\left|\mathbf{u}\right|^2\right)
\\
\left(\frac{\partial}{\partial t}
+\pounds_{\boldsymbol{u}}\right)\boldsymbol{\sigma}&=\mathbf{h}\times\mathbf{n}\,,\hspace{.7cm}
\left(\frac{\partial}{\partial t}
+\pounds_{\boldsymbol{u}}\right)\mathbf{n}=J^{-1}\boldsymbol{\sigma}\times\mathbf{n}.
\label{EL2-bis}
\end{align}
Then, one simply calculates
\begin{align*}
\left(\frac{\partial}{\partial t}
+\pounds_{\boldsymbol{u}}\right)\boldsymbol{\mathcal{C}}_\text{\tiny
EL}
=&
\left(\frac{\partial}{\partial t}
+\pounds_{\boldsymbol{u}}\right)\big(\boldsymbol{u}+\boldsymbol{\sigma}_\text{\tiny
EL}\cdot\mathbf{n}\times\nabla\mathbf{n}\big)
\\
=&\ \nabla\mathbf{n}\cdot\mathbf{h} -
\nabla\left(p+F-\frac12\left|\mathbf{u}\right|^2\right) +
\mathbf{h}\times\mathbf{n}\cdot\mathbf{n}\times\nabla\mathbf{n}
\\
& - J^{-1}\boldsymbol{\sigma}_\text{\tiny
EL}\cdot\nabla\mathbf{n}\times\left(\boldsymbol{\sigma}_\text{\tiny
EL}\times\mathbf{n}\right) + J^{-1}\boldsymbol{\sigma}_\text{\tiny
EL}\cdot\mathbf{n}\times\left(\nabla\boldsymbol{\sigma}_\text{\tiny
EL}\times\mathbf{n}\right)
\\
&+ J^{-1}\boldsymbol{\sigma}_\text{\tiny
EL}\cdot\mathbf{n}\times\left(\boldsymbol{\sigma}_\text{\tiny
EL}\times\nabla\mathbf{n}\right).
\end{align*}
At this point, standard vector identities yield
\begin{align*}
\mathbf{h}\times\mathbf{n}\cdot\mathbf{n}\times\nabla\mathbf{n}&=\left(\mathbf{h}\cdot\mathbf{n}\right)\left(\nabla\mathbf{n}\cdot\mathbf{n}\right)-\left(\nabla\mathbf{n}\cdot\mathbf{h}\right)\left(\mathbf{n}\cdot\mathbf{n}\right)
\\
& = -\nabla\mathbf{n}\cdot\mathbf{h}
\\
\nabla\mathbf{n}\times\left(\boldsymbol{\sigma}_\text{\tiny
EL}\times\mathbf{n}\right)&=\left(\nabla\mathbf{n}\cdot\mathbf{n}\right)\boldsymbol{\sigma}_\text{\tiny
EL}-\left(\nabla\mathbf{n}\cdot\boldsymbol{\sigma}_\text{\tiny
EL}\right)\mathbf{n}
\\
&=-\left(\nabla\mathbf{n}\cdot\boldsymbol{\sigma}_\text{\tiny
EL}\right)\mathbf{n}
\\
\mathbf{n}\times\left(\nabla\boldsymbol{\sigma}_\text{\tiny
EL}\times\mathbf{n}\right)&=\left(\mathbf{n}\cdot\mathbf{n}\right)\nabla\boldsymbol{\sigma}_\text{\tiny
EL}-\left(\nabla\boldsymbol{\sigma}_\text{\tiny
EL}\cdot\mathbf{n}\right)\mathbf{n}
\\
& =\nabla\boldsymbol{\sigma}_\text{\tiny
EL}-\left(\nabla\boldsymbol{\sigma}_\text{\tiny
EL}\cdot\mathbf{n}\right)\mathbf{n}
\\
\mathbf{n}\times\left(\boldsymbol{\sigma}_\text{\tiny
EL}\times\nabla\mathbf{n}\right)&=\left(\nabla\mathbf{n}\cdot\mathbf{n}\right)\boldsymbol{\sigma}_\text{\tiny
EL}-\left(\boldsymbol{\sigma}_\text{\tiny
EL}\cdot\mathbf{n}\right)\nabla\mathbf{n}
\\
&= 0\,,
\end{align*}
where we have made use of the relations $|\mathbf{n}|^2=1$ and
$\boldsymbol{\sigma}_\text{\tiny EL}\cdot\mathbf{n}=0$. Therefore
equation \eqref{Cequation} follows directly from
\[
\left(\frac{\partial}{\partial t}
+\pounds_{\boldsymbol{u}}\right)\boldsymbol{\mathcal{C}}_\text{\tiny
EL}=-\nabla\left(p+F-\frac12\left|\mathbf{u}\right|^2-\frac1{2J}|\boldsymbol{\sigma}_\text{\tiny
EL}|^2\right) \,.
\]
The equation \eqref{EulerForNematics} follows by taking the curl of
the above equation, upon recalling that the curl is given by the
exterior differential, so that
$\mathbf{d}\!\left(\boldsymbol{\mathcal{C}}_\text{\tiny
EL}\cdot\mathrm{d}\mathbf{x}\right)=(\nabla\times\boldsymbol{\mathcal{C}}_\text{\tiny
EL})\cdot\mathrm{d}\mathbf{S}$. Since the differential always
commutes with the Lie derivative \cite{MaRa}, equation
\eqref{EulerForNematics} follows immediately. It is also easy to see
that equation \eqref{h-equation} arises by calculating $
\left(\partial_t
+\pounds_{\boldsymbol{u}}\right)\left(\boldsymbol{\mathcal{C}}_\text{\tiny
EL}\cdot\overline{\boldsymbol{\omega}}_\text{\tiny EL}\right)
=\overline{\boldsymbol{\omega}}_\text{\tiny EL}\cdot\nabla\phi$.

\subsection{Derivation of equations \eqref{Ceq-Qtensor} and \eqref{LdGEuler}\label{appendix2}}
Upon introducing the Lie derivative notation, the equations
\eqref{QS1}-\eqref{QS2}-\eqref{QS3} read as
\begin{align}\label{QS1-bis}
&\left(\frac{\partial}{\partial t} +\pounds_{\boldsymbol{u}}\right)
\boldsymbol{u} = \mathsf{h}_{ij}\nabla\mathsf{Q}_{ij}-\nabla\left(
p+\mathcal{F} -\frac12|\boldsymbol{u}|^2\right)\,,\ \ \nabla\cdot\boldsymbol{u}
=0
\\ \label{QS2-bis}
&\left(\frac{\partial}{\partial t} +\pounds_{\boldsymbol{u}}\right)
\mathsf{Q}=J^{-1}\, \mathsf{P}
\\
&\left(\frac{\partial}{\partial t}
+\pounds_{\boldsymbol{u}}\right)\mathsf{P}=-\mathsf{h} -\lambda
\mathbf{I}
 \label{QS3-bis}
\end{align}
where we have denoted the molecular field by
\[
\mathsf{h}=\frac{\partial\mathcal{F}}{\partial
\mathsf{Q}}-\partial_i\frac{\partial\mathcal{F}}{\partial
\mathsf{Q}_{,i}} \,.
\]
Then, one simply calculates
\begin{multline*}
\left(\frac{\partial}{\partial t}
+\pounds_{\boldsymbol{u}}\right)\boldsymbol{\mathcal{C}}_\text{\tiny
QS} = \left(\frac{\partial}{\partial t}
+\pounds_{\boldsymbol{u}}\right)\Big(\boldsymbol{u}+\mathsf{P}_{ij}\nabla\mathsf{Q}_{ij}\Big)
\\
=\mathsf{h}_{ij}\nabla\mathsf{Q}_{ij}-\nabla\left(
p+\mathcal{F} -\frac12|\boldsymbol{u}|^2\right)-\mathsf{h}_{ij}\nabla\mathsf{Q}_{ij}-\lambda\,\delta_{ij}\nabla\mathsf{Q}_{ij}+\frac{1}{J}\mathsf{P}_{ij}
\nabla \mathsf{P}_{ij} \,,
\end{multline*}
which becomes
\[
\left(\frac{\partial}{\partial t}
+\pounds_{\boldsymbol{u}}\right)\boldsymbol{\mathcal{C}}_\text{\tiny
QS} =
-\nabla\!\left(p+\mathcal{F}+\lambda\,\delta_{ij\,}\mathsf{Q}_{ij}-\frac12|\boldsymbol{u}|^2-\frac{1}{2J}\mathsf{P}_{ij}\mathsf{P}_{ij}\right)
\,.
\]
Finally, taking the curl of the above equation returns
\eqref{LdGEuler}.

\subsection{Derivation of equation \eqref{Ceq-Omodel}\label{appendix3}}
Upon using the Lie derivative notation, equations
\eqref{SG1}-\eqref{SG2}-\eqref{SG3} may be written as
\begin{align}\label{SG1-bis}
&\left(\frac{\partial}{\partial t} +\pounds_{\boldsymbol{u}}\right)
\boldsymbol{u} =-
\sigma_r\nabla\frac{\partial\mathcal{E}}{\partial\sigma_r}+\frac{\partial\mathcal{E}}{\partial
O_{mn}}\nabla O_{mn}
 -\nabla\!\left(p-\frac12\left|\boldsymbol{u}\right|^2\right)
\\ \label{SG2-bis}
&\left(\frac{\partial}{\partial t}
+\pounds_{\boldsymbol{u}}\right)\sigma_r=\varepsilon_{r\!ji}\left(\sigma_i\frac{\partial\mathcal{E}}{\partial
\sigma_j} - O_{ih}\frac{\partial\mathcal{E}}{\partial O_{jh}}\right)
\\
&\left(\frac{\partial}{\partial t} +\pounds_{\boldsymbol{u}}\right)
O_{mn}=-\varepsilon_{mkj}\frac{\partial\mathcal{E}}{\partial
\sigma_j}O_{kn} \label{SG3-bis}
\end{align}
so that, upon denoting $\phi=p-|\boldsymbol{u}|^2/2$, one computes
\begin{align*}
\left(\frac{\partial}{\partial t}
+\pounds_{\boldsymbol{u}}\right)\boldsymbol{\mathcal{C}}
 = & \left(\frac{\partial}{\partial t}
+\pounds_{\boldsymbol{u}}\right)\left(\boldsymbol{u}+\frac12\,\varepsilon_{mns}\,\sigma_s\,
O_{mk}\nabla O_{nk}\right)
\\
=&-
\sigma_r\nabla\frac{\partial\mathcal{E}}{\partial\sigma_r}+\frac{\partial\mathcal{E}}{\partial
O_{mn}}\nabla O_{mn}
 -\nabla\phi
 \\
 &+
 \frac12\varepsilon_{mns}\left(\varepsilon_{sji}\,\sigma_i\frac{\partial\mathcal{E}}{\partial
\sigma_j}-\varepsilon_{sji}\,O_{ih}\frac{\partial\mathcal{E}}{\partial
O_{jh}}\right)O_{mk}\nabla O_{nk}
\\
 &-
 \frac12\varepsilon_{mns}\,\sigma_s\,\varepsilon_{mhj}\,\frac{\partial\mathcal{E}}{\partial
\sigma_j}O_{hk}\nabla O_{nk}
\\
 &-
 \frac12\varepsilon_{mns}\,\sigma_s\,O_{mk}\,\varepsilon_{nhj}\left(\nabla\frac{\partial\mathcal{E}}{\partial
\sigma_j}\,O_{hk}+\frac{\partial\mathcal{E}}{\partial
\sigma_j}\nabla O_{hk}\right) .
\end{align*}
At this point we observe that, since
$O_{mk}O_{hk}=O_{mk}O_{kh}^{-1}=\delta_{mh}$, then
\begin{align*}
-
\sigma_r\nabla\frac{\partial\mathcal{E}}{\partial\sigma_r}-\frac12\,\varepsilon_{mns}\,\sigma_s\,O_{mk}\,\varepsilon_{nhj}\nabla\frac{\partial\mathcal{E}}{\partial
\sigma_j}\,O_{hk}=& -
\sigma_r\nabla\frac{\partial\mathcal{E}}{\partial\sigma_r}-\frac12\,\varepsilon_{hns}\,\varepsilon_{nhj}\,\sigma_s\nabla\frac{\partial\mathcal{E}}{\partial
\sigma_j}
\\
&= -
\sigma_r\nabla\frac{\partial\mathcal{E}}{\partial\sigma_r}+\delta_{sj}\,\sigma_s\nabla\frac{\partial\mathcal{E}}{\partial
\sigma_j}=0.
\end{align*}
Moreover, the sum of all terms in
$\partial\mathcal{E}/\partial\boldsymbol{\sigma}$ can be written as
\begin{align*}
&\sigma_s\,\frac{\partial\mathcal{E}}{\partial\sigma_j}\left(\varepsilon_{ijs}\,\varepsilon_{mni}\,O_{mk}\nabla
O_{nk}-\varepsilon_{ims}\,\varepsilon_{inj}\,O_{nk}\nabla
O_{mk}-\varepsilon_{mis}\,\varepsilon_{inj}\nabla
O_{nk}\,O_{mk}\right)
\\
= &\
-\sigma_s\,\frac{\partial\mathcal{E}}{\partial\sigma_j}\,\varepsilon_{ijs}\,\varepsilon_{imn}\,O_{mk}\nabla
O_{nk}
\\
=& \
-\sigma_s\,\frac{\partial\mathcal{E}}{\partial\sigma_j}\left(\delta_{jm}\,\delta_{sn}-\delta_{jn}\,\delta_{sm}\right)O_{mk}\nabla
O_{nk}
\\
=&\
-\sigma_s\,\frac{\partial\mathcal{E}}{\partial\sigma_j}\left(\nabla
O_{sk}\,O_{kj}^{-1}+O_{sk}\nabla O_{kj}^{-1}\right)
\\
=&-\sigma_s\,\frac{\partial\mathcal{E}}{\partial\sigma_j}\,\nabla\!\left(O_{sk}\,O_{kj}^{-1}\right)=0.
\end{align*}
In addition, we calculate
\begin{align*}
&\frac{\partial\mathcal{E}}{\partial O_{mn}}\,\nabla
O_{mn}-\frac12\,\varepsilon_{sji}\,\varepsilon_{mns}\,O_{ih}\,\frac{\partial\mathcal{E}}{\partial
O_{jh}}\,O_{mk}\,\nabla O_{nk}
\\
=&\frac{\partial\mathcal{E}}{\partial O_{mn}}\,\nabla
O_{mn}+\frac12\left(\delta_{mj}\,\delta_{ni}-\delta_{mi}\,\delta_{nj}\right)O_{ih}\,\frac{\partial\mathcal{E}}{\partial
O_{jh}}\,O_{mk}\,\nabla O_{nk}
\\
=& \frac{\partial\mathcal{E}}{\partial O_{mn}}\,\nabla
O_{mn}+\frac12\left(O_{ih}\,O_{jk}\,\nabla
O_{ik}-O_{ih}\,O_{ik}\,\nabla
O_{jk}\right)\frac{\partial\mathcal{E}}{\partial O_{jh}}
\\
=& \frac{\partial\mathcal{E}}{\partial O_{mn}}\,\nabla
O_{mn}+\frac12\left(O_{hi}^{-1}\,\nabla
O_{ik}\,O_{kj}^{-1}-\delta_{hk}\,\nabla
O_{jk}\right)\frac{\partial\mathcal{E}}{\partial O_{jh}}
\\
=& \frac{\partial\mathcal{E}}{\partial O_{mn}}\,\nabla
O_{mn}-\frac12\left(\nabla O_{hj}^{-1}+\nabla
O_{jh}\right)\frac{\partial\mathcal{E}}{\partial O_{jh}}
\\
=& \frac{\partial\mathcal{E}}{\partial O_{mn}}\,\nabla O_{mn}-\nabla
O_{jh}\frac{\partial\mathcal{E}}{\partial O_{jh}}=0.
\end{align*}
Thus, we have proved the relation
\begin{align*}
\left(\frac{\partial}{\partial t}
+\pounds_{\boldsymbol{u}}\right)\boldsymbol{\mathcal{C}}=-\nabla\phi
\,,
\end{align*}
whose curl yields the corresponding Euler-like equation
\eqref{NewCequation}, thereby recovering the corresponding helicity
conservation. Notice that the above result also holds in the case of
explicit dependence of the free energy $\mathcal{E}$ on the gradient
$\nabla O$ of the orientational order parameter.


\end{document}